\documentstyle[epsfig,12pt]{article}
\def\avg#1{\langle{#1}\rangle}
\def\dd{_{\!_\Delta}}

\begin{document}

\parindent=0cm
\baselineskip=16pt

\title{Efficiency in foreign exchange markets}
\author{
Roberto~Baviera$^{1}$,
Michele~Pasquini$^{2}$, 
Maurizio~Serva$^{3}$, \\
Davide~Vergni$^{4}$ \&
Angelo~Vulpiani$^{5}$ \\}
\maketitle
\vspace*{2truecm}

{\small \it
\begin{tabular}{ll}
$1$ & Dipartimento di Fisica, Universit\`a dell'Aquila, 
      \& I.N.F.M., Unit\`a dell'Aquila \\
    & I-67100 Coppito, L'Aquila, Italy.  E-mail : {\rm baviera@axtnt1.roma1.infn.it} 
      \vspace*{0.2truecm} \\

$2$ & Dipartimento di Matematica, Universit\`a dell'Aquila, 
      \& I.N.F.M., Unit\`a dell'Aquila \\
    & I-67100 Coppito, L'Aquila, Italy.  E-mail : {\rm pasquini@serva.dm.univaq.it}
      \vspace*{0.2truecm} \\

$3$ & Dipartimento di Matematica, Universit\`a dell'Aquila, 
      \& I.N.F.M., Unit\`a dell'Aquila \\
    & I-67100 Coppito, L'Aquila, Italy.  E-mail : {\rm maurizio.serva@aquila.infn.it}
      \vspace*{0.2truecm} \\

$4$ & Dipartimento di Fisica, Universit\`a di Roma ``La Sapienza''
      \& I.N.F.M., Unit\`a di Roma 1 \\
    & I-00185 P.le A.~Moro 2, Roma, Italy.  E-mail {\rm davide.vergni@roma1.infn.it}
      \vspace*{0.2truecm} \\

$5$ & Dipartimento di Fisica, Universit\`a di Roma ``La Sapienza''
      \& I.N.F.M., Unit\`a di Roma 1 \\
    & I-00185 P.le A.~Moro 2, Roma, Italy.  E-mail {\rm angelo.vulpiani@roma1.infn.it} \\
\end{tabular}
}

\vspace{3.0cm}
\begin{flushleft}
{\bf Address for correspondence:}\\
Angelo Vulpiani\\
Dipartimento di Fisica\\
Universit\`a di Roma ``La Sapienza''\\
I-00185 Piazzale Aldo Moro 2, Roma, Italy\\
Tel. +39-06-4991 38 40\\
Fax  +39-06-446 31 58 \\
angelo.vulpiani@roma1.infn.it
\end{flushleft}

\newpage
$\:\:$
\newpage

\begin{center}
{\bf\Large
Efficiency in foreign exchange markets}
\end{center}

\vspace{0.5truecm}

\begin {abstract}
In this paper we test the efficiency hypothesis in financial market.
A market is called efficient if the price variations ``fully reflect''
relevant information, i.e. a speculator cannot make a profit
out of it.
A currency exchange market is a natural candidate to check efficiency 
because of its high liquidity.
We perform a statistical study  of {\it weak} efficiency in
Deutschemark/US dollar exchange rates using high frequency data.
In the {\it weak} form of efficiency the information can only
come from historical prices.

The presence of correlations in the returns sequence 
implies the possibility of a statistical prevision of market behavior. 
We show the existence of correlations by means two statistical tools.
A first analysis has been performed using structure functions.
This approach gives an indication on the returns distributions
at different lags $\tau$.
We have also computed the generalized correlation functions of the return
absolute values; roughly speaking this is a test of the independence
of the fluctuations of fixed size.
In both cases we have obtained a clear evidence of long term return anomalies. 
This implies a failure of the usual ``random walk'' model of the returns;
nevertheless the presence of long term correlations does not 
directly imply the fault of the {\it weak} efficiency hypothesis~:
it is not obvious how to use time correlation to make
a profit in a realistic investment.

Then we show how this information is relevant for a speculator.
First we introduce a measure of the {\it available} information
relevant from a financial point of view, with a technique which
reminds the Kolmogorov $\epsilon$-entropy.
Second in the case of no transaction costs, we propose a simple 
investment strategy which leads to an exponential growth rate 
of the capital related to the {\it available} information. 

We have performed two kind of information analysis in the return series.
We show that the {\it available} information is practically zero
if the speculator wants to change his portfolio {\it systematically}
after a fixed lag $\tau$~: for him the market is efficient.
Instead, a finite {\it available} information is observed by
a {\it patient} investor who cares only of fluctuation
of given size $\Delta$.
This is the first case, as far as we know,
in which the {\it available} information
obtained by a suitable data analysis
is directly linked to the possible earnings
of a speculator who follows a particular trading rule.

\end{abstract}

\vspace{1.0truecm}

\section{Introduction}

A large amount of research suggests that prices are related 
with information, and in particular it focuses on 
efficiency in financial markets.
A market is inefficient if a speculator
can make a profit out of information present in the market.
Since the celebrated work of Fama~\cite{FamaI}
a big effort has been done to test empirically and
to understand theoretically the efficiency of financial
markets.

A market is said to be efficient if prices ``fully reflect''
all available information, i.e. such information is completely exploited 
in order to determine the price, after having taken into account
the costs to use this information and
a transient time, due to costs, to reach equilibrium. 
The idea is that the investor destroys information while using it
and as a consequence he contributes to produce equilibrium.

In the last years long term correlations have been observed
in financial markets. 
We shall not review in details the contributions to the field. 
We stress that long term return anomalies   
are usually revealed via test of efficiency in a {\it semi-strong} form,
i.e. not only considering the asset prices but also 
some other publicly known news.  
The interest is generally focused on the market reactions 
to an event occurred a fixed period time before (three to five typically)
such as divested firms~\cite{Cusatis}, mergers~\cite{Asquith} 
or initial public offerings~\cite{RitterI,RitterII}.
Recent research~\cite{DGE,BB,DLC,Pagan,paserva1,paserva2} 
has pointed out the existence of long range correlations
also in the {\it weak} form.
However only low frequency data are considered 
and implications on efficiency are not completely understood.

In this paper we focus on efficiency in the {\it weak} form, 
i.e we consider only the information coming from historical prices.
We are interested on a time scale longer than the typical
correlation returns time (few minutes)
but lower than the characteristic time after which
we do not have statistical relevance of the results:
in this sense we deal with {\it long term} return anomalies. 
Currency exchange seems to be the natural subject for an efficiency test.
We expect that such markets are very efficient
as a consequence of the large liquidity.

For these reasons we have decided to 
analyze a one year high frequency dataset
of the Deutschemark/US dollar exchange,
the most liquid market.
Our data, made available by Olsen and Associated,
contains all worldwide $1,472,241$ bid--ask Deutschemark/US dollar
exchange rate quotes registered by the inter-bank Reuters network 
over the period October 1, 1992 to September 30, 1993. 

One of the main problem in tick data analysis,
is the irregular spacing of quotes.
In this paper we consider {\it business} time,
i.e. the time of the transaction given by its rank 
in the sequence of quotes.
This seems to be a reasonable way to consider time 
in a worldwide time series, where time delays and lags
of no transaction are often due to geographical reasons.

In this paper we test the independence hypothesis of returns and 
define and measure an {\it available} information.
In section {\bf 2} we check the independence with two 
different techniques.
The first one, called structure functions analysis, shows whether
it is possible to rescale properly the distribution functions 
at different lags~\cite{Mandelbrot}. 
The second one is a direct independence test.
The independence of two random variables $x$, $y$ implies that $f(x)$ and
$g(y)$ are uncorrelated for every $f$ and $g$.
We check it for $f(\cdot)=g(\cdot)=|\cdot|^q$.
We interpret these quantities as an estimate of the correlation between
returns of given size.

We want to quantify the {\it available} information
and discuss its financial relevance.
In section {\bf 3} we consider a speculator with a given resolution,
i.e. he is concerned only about fluctuations at least of size $\Delta$.
This reminds the $\epsilon$ entropy introduced by 
Kolmogorov~\cite{Kolmogorov} in the context of information theory.
A similar filter has been first introduced by
Alexander~\cite{AlexanderI,AlexanderII}.
To show the inefficiency of the market he proposed the following trading rule~: 
if the return moves up of $\Delta$, buy and hold  until it goes down 
of $\Delta$ from a subsequent high, then sell and maintain the short
position till the return rises again of $\Delta$ above a subsequent low.

Here we divide the problem in two parts.
First we define the {\it available} information for any fixed
resolution $\Delta$ of the speculator.
Second, as suggested by Fama~\cite{FamaI}, we relate the {\it available}
information with the profitability by means of a particular trading rule.
We show how this information is related to the optimal growth rate portfolio
using a simple approximation in terms of Markov process.

In section {\bf 4} we summarize and discuss the results.

\section{Long term correlations}

After the seminal work of Bachelier~\cite{Bachelier},
it was widely believed that the price variations 
follow an independent, zero mean, gaussian process.
The main implications of the ``fundamental principle'' of Bachelier  
are that the price variation is a martingale and 
it is an independent random process.

Bachelier considers the market a ``fair game''~:
a speculator cannot exploit previous information to make
better predictions of forthcoming events.
Information can come only from correlations and 
in absence of them from the shape of the probability 
distribution of the returns.

For about sixty years this contribution was practically forgotten, 
and quantitative analysis
on financial data started again with advent of computers.

Following Fama~\cite{FamaI}, we shall call hereafter ``random walk''
the financial models where the returns
 
\begin{equation} 
r_t \equiv \ln \frac{S_{t+1}}{S_t} 
\label{eq:r_t-definition}
\end{equation} 
are independent variables. 
In this paper we define $S_t$ as the average between bid and ask price.
We do not want to enter here in a detailed analysis of
the huge literature about ``random walk'' models.
We just mention that,
before the contribution of Mandelbrot~\cite{MandelbrotII},
the return $r_t$
was considered well approximated by an independent gaussian process.
Mandelbrot proposed that the returns were 
distributed according a Levy-stable,
still remaining independent random variables.  

At present, it is commonly accepted that the variables
\begin{equation} 
r^{(\tau)}_t \equiv \sum^{t+\tau}_{t'=t+1} r_{t'} =\ln \frac{S_{t+\tau}}{S_t}
\label{eq:r_tau-definition}
\end{equation} 
do not behave according a gaussian at small $\tau$,
while the gaussian behavior is recovered for large $\tau$.
Of course a return $r_t$ distributed according to a Levy,
as suggested by Mandelbrot, is stable under composition 
and then also $r^{(\tau)}_t$ would follow the same distribution for every $\tau$.
A recent proposal is the truncated Levy distribution model introduced by  
Mantegna and Stanley~\cite{MantegnaStanley}
which fits well the data and reproduces the transition
from small to large $\tau$.

Let us focus our attention on independence tests.
We remark once again that 
an influence of the return $r_t$ at time $t$ on 
the return $r_{t+\tau}$ at time $t+\tau$
implies a not fully efficient
market in a {\it weak} form.
The relevance of the question is clear in the case of an investor
analyzing historical data to a make market forecast and a profit
out of it.

As a test of independence it is generally considered the correlation
functions on time intervals $\tau$
\begin{equation} 
C(\tau) \equiv \avg{r_t r_{t+\tau}}-\avg{r_t} \avg{r_{t+\tau}}\,\,,
\label{eq:correlation}
\end{equation}
where $\avg{\cdot}$ denotes the temporal average 
$$
\avg{A} \equiv \frac{1}{T} \sum^{T}_{t=1} A_t
$$
and $T$ is the size of the sample.

The presence of correlations in Deutschemark/US dollar exchange returns 
before the nineties is a well known fact.
For example in ~\cite{DeJong}, where it is considered the same dataset we use, 
it is shown that the returns are negatively correlated for about three minutes.

We remind that in general uncorrelation does not 
imply independence. 
A sort of long term memory
can be revealed with appropriate tools,
see for example the seminal
works in the field of Alexander~\cite{AlexanderI,AlexanderII} and
Niederhoffer and Osborne~\cite{Niederhoffer},
and the most recent literature
\cite{DGE,BB,DLC,Pagan,Taylor,Baillie},
where it is shown that absolute returns or
powers of returns exhibit a long range correlation.
It is a common belief that it is not possible to exploit  
this kind of information because of transaction costs.

We shall show in next section that dependent 
(even if uncorrelated) returns
have a clear financial meaning because they
imply the existence of {\it available} information. 

In subsection {\bf 2.1} we show the persistence
of a long range memory for the Deutschemark/US dollar exchange rate
by means of the analysis of structure functions.
In subsection {\bf 2.2}, we test directly the independence 
of returns with a generalization of the 
correlation analysis.

\subsection{Structure functions}

There is some evidence that the process
$r^{(\tau)}_t$ cannot be described in terms of a 
unique scaling exponent~\cite{Ghashghaie,Vattay},
i.e. it is not possible to find a real number $h$ such that
the statistical properties of the new random variable $r^{(\tau)}_t/\tau^h$
do not depend on $\tau$.

The scaling exponent $h$ gives us information on the features 
of the underlying process.
In the case of independent gaussian behavior of $r_t$
the scaling exponent is $1/2$.

On the contrary, the data show that the probability distribution
function of $r^{(\tau)}_t/\sqrt{Var[r^{(\tau)}_t]}$ changes with 
$\tau$~\cite{Ghashghaie,Vattay}.
This is an indication 
that $r_t$ is a dependent stochastic process
and it implies the presence of wild fluctuations.

A way to show these features, which is standard for the fully developed turbulence
theory~\cite{Frisch}, is to study the structure functions~:
\begin{equation} 
F_q(\tau) \equiv \avg{|r^{(\tau)}_t|^q} \,\,.
\label{eq:structure_functions}
\end{equation} 

In the simple case where 
$r_t$ is an independent random process, one has (for a certain range of $\tau$)
\begin{equation} 
F_q(\tau) \sim \tau^{hq} \,\,,
\label{eq:selfaffine}
\end{equation} 
where $h<1/2$ in the Levy-stable case while the gaussian behavior is
recovered for $h=1/2$.
The truncated Levy distribution corresponds to $h<1/2$ for $\tau$
sufficiently small 
and to $h=1/2$ at large $\tau$.
``Random walk'' models present always a unique scaling exponent. 
If the structure function has the behavior in (\ref{eq:selfaffine}) 
we call the process self-affine (sometimes called uni-fractal). 

\begin{figure}[htb]
 \begin{center}
  \resizebox{0.8\textwidth}{!}{
     \includegraphics{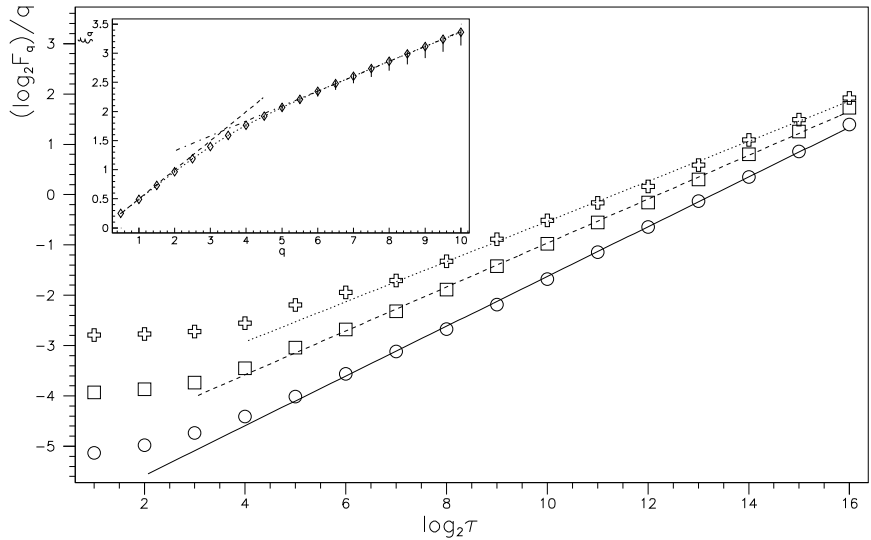}
  }
  \protect\caption{Structure functions $\frac{1}{q}\log_2 F_q(\tau)$ 
	           {\it versus} $\log_2 \tau$ for Deutschemark/US 
		   dollar exchange rate quotes. The three plots
		   correspond to different value of $q$~: $q = 2.0$ ($\circ$),
		   $q = 4.0$ ($\Box$) and $q = 6.0$ ($+$). In the insert we show
		   $\xi_q$ versus $q$. We estimate with linear regression two different
		   regions in this graph. The first one is a line of slope $0.5$ (dashed line),
		   and the second has a slope $0.256$ (dash dotted line).}
  \label{fig:Fni Strutt}
 \end{center}
\end{figure}

As previously mentioned a description in terms of a unique scaling exponent
$h$, does not work.
Therefore instead of (\ref{eq:selfaffine}) one has 
\begin{equation} 
F_q(\tau) \sim \tau^{\xi_{q}} \,\,,
\label{eq:multiaffine}
\end{equation}
where $\xi_{q}$ are called scaling exponents of order $q$.
If $\xi_{q}$ is not linear, the process is called multi-affine
(sometimes multi-fractal). 
Using simple arguments it is possible to see that $\xi_{q}$ 
has to be a convex function of $q$~\cite{Feller}.
The larger is the difference of $\xi_{q}$ from the linear behavior
in $q$ the wilder are the fluctuations and the correlations of returns.
In this sense the deviation from a linear shape for $\xi_q$ gives 
an indication of the relevance of correlations.

In figure~\ref{fig:Fni Strutt} we plot,
the $F_q(\tau)$ for three different
values of $q$.
A multi-affine behavior 
is exhibited by different slopes of $\frac{1}{q} \log_2 (F_q)$
{\it vs.} $\log_2 (\tau)$,
at least for $\tau$ between $2^4$ and $2^{15}$.
For larger {\it business} lags a spurious behavior can arise
because of the finite size of the dataset considered.
In the insert we plot the $\xi_q$ estimated 
by standard linear regression 
of $\log_2 F_q(\tau)$ {\it vs.} $\log_2 (\tau)$
for the values of $\tau$
mentioned before.
To give an estimation of errors, 
the most natural way turns out to be a division
of the year dataset in two semesters.
This is natural in the financial context,
since it is a measure of reliability of the second semester
forecast based on the first one.
We observe that the traditional stock market theory
(brownian motion for returns), 
gives a reasonable agreement with $\xi_q \simeq q/2$ only for $q<3$,
while for $q>6$ one as $\xi_q \simeq \tilde{h}q + b$
with $\tilde{h} = 0.256$ and $c = 0.811$.
We stress once again that such a behavior cannot be explained
by a ``random walk'' model (or other self-affine models)
and this effect is a clear evidence 
of correlations present in the signal.

\subsection{Long term correlations analysis}

Let us consider the absolute returns series $\{|r_t|\}$,
which is often shown to be long range correlated in recent literature
\cite{DGE,BB,DLC,Pagan,paserva1,paserva2,Taylor,Baillie}.
Absolute values mean that we are interested only in the
size of fluctuations.

Let us introduce the generalized correlations $C_q(\tau)$:
\begin{equation}
C_q(\tau) \equiv \avg{|r_t|^q |r_{t+\tau}|^q} - 
		 \avg{|r_t|^q} \avg{|r_{t+\tau}|^q} \,\,.
\end{equation}
We shall see that the above functions will be 
a powerful tool to study correlations of returns
with comparable size:
small returns are more relevant at small $q$,
while $C_q(\tau)$ is dominated by large returns at large $q$
(the usual definition of correlation 
for absolute returns
is recovered for $q=1$).

Following the definitions in \cite{BD}, 
let us suppose to have a long memory for the absolute returns series,
i.e. the correlations $C_q(\tau)$ approaches zero very slowly 
at increasing $\tau$, i.e. $C_q(\tau)$ is a power-law:
$$
C_q(\tau) \sim \tau^{-\beta_q} \,\, . 
$$

If $|r_t|^q$ is an uncorrelated process one has $\beta_q = 1$,
while $\beta_q$ less than $1$ corresponds to long range memory.

Instead of directly computing correlations $C_q(\tau)$ of single returns 
we consider rescaled sums of returns.
This is a well established way,
if one is interested only in long term analysis, 
in order to drastically reduce
statistical errors that can affect our quantities~\cite{Papoulis}.
Let us introduce the {\it generalized cumulative absolute returns}
\cite{paserva1,paserva2}
\begin{equation}
   \chi_{t,q}(\tau) \equiv \frac{1}{\tau} \sum_{i=0}^{\tau-1} |{r_{t+i}}|^q
   \label{eq:defchi}
\end{equation}
and their variance
\begin{equation}
   \delta_q(\tau) \equiv 
	\avg{\chi_{t,q}(\tau)^2} - {\avg{ \chi_{t,q}(\tau)}}^2 \,\,.
   \label{eq:defvarchi}
\end{equation}
After some algebra (see Appendix), one can show
that if $C_q(\tau)$ for large $\tau$ 
is a power-law with exponent $\beta_q$, then
$\delta_q(\tau)$ is a power-law with the same exponent~:
$$
C_q(\tau) \sim \tau^{-\beta_q} \ \ \
\Longrightarrow  \ \ \
\delta_q(\tau) \sim \tau^{-\beta_q} \,\,.
$$

In other words the hypothesis of long range memory 
for absolute returns ($\beta_q < 1$),
can be checked via the numerical analysis of $\delta_q(\tau)$.

\begin{figure}[hbt]
 \begin{center}
  \resizebox{0.8\textwidth}{!}{
     \includegraphics{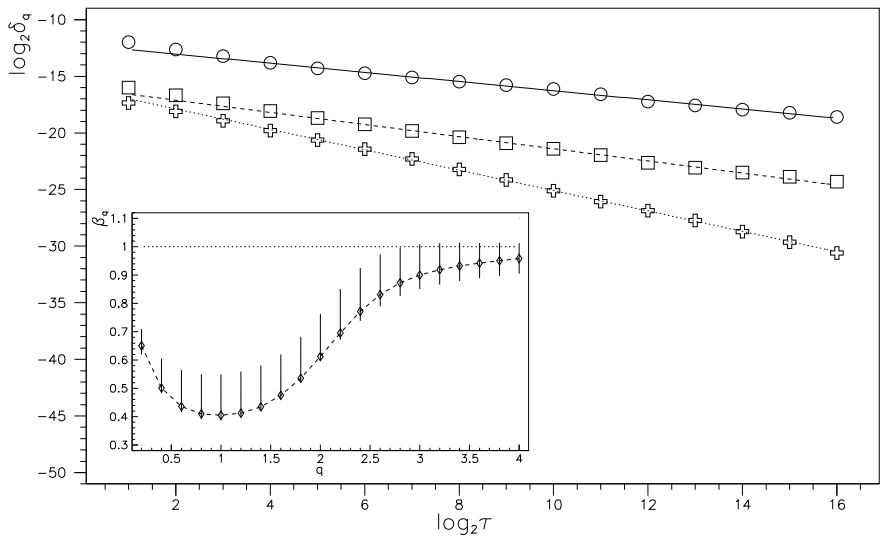}
  }
  \protect\caption{$\log_2 \delta_q$ {\it versus} $\log_2 \tau$. The three plots
		   correspond to different value of $q$~: $q = 1.0$ ($\circ$),
		   $q = 1.8$ ($\Box$) and $q = 3.0$ ($+$). In the insert we show
		   $\beta_q$ versus $q$, the horizontal line shows value $\beta = 1$
		   corresponding to independent variable.}
  \label{fig:correlations}
 \end{center}
\end{figure}

In figure~\ref{fig:correlations} we plot the $\delta_q$ {\it vs.} $\tau$
in log-log scale, for three different values of $q$.
The variance $\delta_q(\tau)$ is affected
by small statistical errors, and it confirms 
the persistence of a long range memory for a $\tau$ larger than 
$2^4$ and up to $2^{15}$.

The exponent $\beta_q$ can be profitably estimated 
by standard linear regression of 
$\log_2(\delta_q(\tau))$ {\it versus} $\log_2(\tau)$,
and the errors are estimated in the same way of subsection {\bf 2.1}.

We notice in the insert that the ``random walk'' model behavior
is remarkably different from the one observed in the
Deutschemark/US dollar exchange for $q < 3$.
This implies the presence of strong correlations,
while one has $\beta_q=1$ for large values of $q$,
i.e. big fluctuations are practically independent.

An intuitive meaning of the previous results is the following.
Using different $q$ one selects different sizes of the fluctuations.
Therefore the non trivial shape of $\beta_q$ is an indication
of the existence of long term anomalies.

\section{Available information}

Let us focus our attention on information analysis of the return $r_t$.
We must treat the dataset in such a way that methods of
information theory can be applied.

The usual approach is the codification of
the original data in a symbolic sequence. 
There are several ways to build up such a sequence: 
one should make sure that this treatment 
does not change the structure of the process
underlying the evolution of the financial data. 

In order to construct a symbolic sequence from a time series, 
at least two steps are needed~:
\begin{itemize}
   \item A {\em filtering} procedure to remove most of the noise
	 in the dataset.
   \item A {\em coarse graining} procedure to partition the range 
	 of variability of the filtered data, 
	 in order to assign a conventional symbol 
         to each element of the partition.
\end{itemize}
The codification is then straightforward:
a symbol corresponds unambiguously to the data stored
in each element of the partition.

From the original signal $r_t$ we obtain a discrete symbolic sequence~:
$$ c_1, c_2, \ldots, c_i, \ldots $$
where each $c_i$ takes only a finite number, say  $m$, of values.
In such a way we reduce ourself to the study of a discrete stochastic process.

A simple way to obtain a symbolic sequence is to consider only
a two-valued symbol and define a discrete random variable without
performing any filtering operation~:
\begin{equation}
   c_i = \left \{ {\begin {array} {ccc}
			  -1 & {\mathrm if} & r_i < 0 \\
			  +1 & {\mathrm if} & r_i \geq 0
		   \end{array} } \right . \,\,\,\,.
   \label{eq:filtro1}
\end{equation}

The financial meaning of this codification is rather evident:
the symbol $-1$ occurs if the stock price decreases, 
otherwise the symbol is $1$.

Let us now remind some basic concepts of information theory.
Consider a sequence of $n$ symbols $C_n=\{c_1,c_2,\ldots,c_n\}$ and its
probability $p(C_n)$.
The block entropy $H_n$ is defined by
\begin{equation}
   H_n \equiv -\sum_{C_n} p(C_n)\ln p(C_n) \,\,.
   \label{def:blockentropy}
\end{equation}

The difference
\begin{equation}
   h_n \equiv H_{n + 1} - H_n
   \label{def:shannonentropy}
\end{equation}
represents the average information needed to specify the symbol $c_{n+1}$ given
the previous knowledge of the sequence $\{c_1,c_2,\ldots,c_n\}$.

The series of $h_n$ is monotonically not increasing and for an {\em ergodic} 
process one has 
\begin{equation}
   h = \lim_{n \rightarrow \infty} h_n
\end{equation}
where $h$ is the Shannon entropy~\cite{Shannon}.

It is easy to show that if the stochastic process $\{c_1,c_2, \ldots \}$
is markovian of order $k$ ({\em i.e.} the probability to have $c_n$
at time $n$ depends only on the previous $k$ steps
$n-1, n-2, \ldots, n - k$), then $h_n = h$ for $n \geq k$. 
In other cases, or $h_n$ goes to zero for increasing $n$ which means 
that either for $n$ sufficiently large the $(n+1)$th-symbol is predictable 
knowing the sequence $C_n$ or it tends to a positive finite value.
The maximum value of $h$ is $\ln(m)$.
It occurs if the process has no memory at all and the $m$
symbols have the same probability.

The difference between $\ln(m)$ and $h$
is intuitively the quantity of information we may use
to predict the next result of the phenomenon we observe,
i.e. the market behavior.
We define {\it available} information~:
\begin{equation}
   I \equiv \ln(m) - h = R \ln(m)
   \label{eq:avainfo}
\end{equation}
where $R=1-h/\ln(m)$ is called the {\it redundancy} of the 
process~\cite{Shannon}.

Hereafter we limit the discrete process to take only two values,
$-1$ and $1$ which have an evident financial meaning.
We expect
that the high frequency details are not relevant and cannot be easily 
used by financial analysts.
It seems rather reasonable to study the process $r_t$ 
with a finite lag $\tau$ (see subsection {\bf 3.1}) or
a finite resolution $\Delta$ on the values of $r_t$ 
(see subsection {\bf 3.2}).

In the following we shall show that different 
discretization procedures
lead to completely different results. 
This corresponds to two different kind of investment,
one {\it systematic} and the other {\it patient}.
The {\it systematic} investor modifies his portfolio every $\tau$ steps,
where the lag $\tau$ is measured in the usual {\it business} time
(but the same results hold also for the {\it calendar} time).
The {\it patient} investor, instead, waits to update his strategy until 
a certain behavior of the market is achieved,
for example, a fluctuation of size $\Delta$.

In the last part of this section we shall show that
this kind of investment seems to be the most suitable 
for financial aims.

\subsection {A naive approach: fixed lag analysis}

In a recently proposed time series model~\cite{Hasbrouck}
the price variation is considered as a result of a 
true underlying process plus an uncorrelated white noise.
If we think the observed return as the sum of these two components,
it is natural to try to eliminate the additional noise
taking the average of the signal over a given lag.

More precisely we treat the financial data as follows~:
\begin{itemize}
   \item we group the sequence of the $r_t$ in 
	 non overlapping blocks of $\tau$ data and we define the
	 sum of the data in the $j^{th}$ block 
	 $$r'_j \equiv \sum_{k = j \tau + 1}^{j(\tau + 1)}r_k \,\, .$$
	 Notice that $r'_j$ is equivalent to $r^{(\tau)}_{j\tau}$ 
	 where $r^{(\tau)}_t$ is defined in 
	 equation~(\ref{eq:r_tau-definition}).
   \item we decimate the data, i.e. we take from the sequence 
	 $r'_j$ only one data every $m$~: $$R_k \equiv r'_{mk} \,\, .$$
	 This procedure should eliminate the eventual short
	 time correlation of the noise.
   \item we build up the symbolic sequence, like in 
	 equation~(\ref{eq:filtro1})~:
	 \begin{equation}
	   c_k = \left \{ {\begin {array} {ccc}
				  -1 & {\mathrm if} & R_k < 0 \\
				  +1 & {\mathrm if} & R_k \geq 0
			   \end{array} } \right . \,\,\,\,.
	   \label{eq:filtrolineare}
	 \end{equation}
\end{itemize}

Let us remark that the first two steps have been performed to reduce the noise.

The total number of the data of the symbolic sequence 
is $N/(m\tau)$, where $N$ is the number of original data.
This filter is linear, i.e. if the signal 
is a linear combination of various contributes, at the end of the filtering
procedure we have the sum of the filtered contributes.
The theory of linear filter is well developed in literature
(see for example \cite{Papoulis}), and we use this simple
approach to check whether a noise is added to our signal.

At this point we have a binary sequence from which we
compute the Shannon entropy.

\begin{figure}[htb]
 \begin{center}
  \resizebox{0.8\textwidth}{!}{\includegraphics{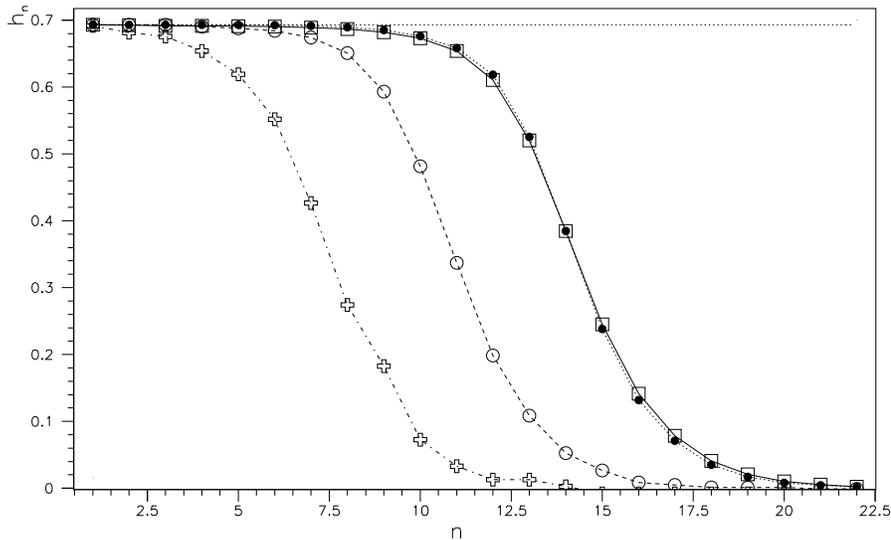}}
  \protect\caption{$h_n$ {\it versus} $n$. The three plots correspond 
		   to different value of $\tau$ and $m$~: $\tau = 10,\,\,m=10$ ($\Box$),
		   $\tau = 10,\,\,m=100$ ($\circ$) and $\tau = 100,\,\,m=100$ ($+$).
		   We show also the entropy numerically obtained from a 
		   coin tossing sequence with the same number of data 
		   of the case $\tau = 10,\,\,m=10$ ($\bullet$). The dotted line
		   indicates $\ln(2)$.}
  \label{fig:entrdecimediate}
 \end{center}
\end{figure}

\noindent Figure \ref{fig:entrdecimediate} reports the results of our analysis
for the linear filter.
We plot $h_n$ vs $n$ for various $\tau$ and $m$,  
compared with the entropy of Bernoulli trials with probability 
$p=0.5$ (this is nothing that the usual coin tossing). 

We know that the entropy $h_n$ of a fair binary Bernoulli trial
must be $\ln(2)$ for every $n$.
The folding of $h_n$ at large $n$ depends
on the finite number of sequence elements. 
It can be proved \cite{Khinchin} that the statistical analysis 
does not give the proper value of $h_n$ for $n$ larger than~:
$$ n^* \approx \frac{1}{h} \ln(N) $$
where $h$ is the entropy of the signal and $N$ is the length
of the sequence.

It should be now clear that the entropy of the sequence
is given by the value of the {\it plateau}.

The entropy does not differ sensibly from $\ln(2)$, of the coin tossing,
and, therefore,  we cannot make prevision on the market.
In conclusion, the financial data cannot 
be represented as a white noise added on a true underlying signal.

Nevertheless, because of the long term correlations (see section {\bf2}),
there is a clear indication that the present state of the market 
depends non trivially on the past.

\subsection {A fixed resolution analysis}

The failure of the previous analysis lead us to try another 
approach in order to keep the information present in the financial data,
this time we use a non-linear filter with a clear financial meaning.

The procedure to create the symbolic sequence is now~:
\begin{itemize}
   \item we fix a resolution value $\Delta$ and we define 
	 \begin{equation}
	    r_{t,t_0} \equiv \ln \frac{S_t}{S_{t_0}} \,\,,
	    \label{eq:rttstar}
	 \end{equation}
	 where $t_0$ is the initial {\it business} time,
	 and $t > t_0$.
	 We wait until an exit time $t_1$ such as~:
	 $$ |r_{t_1,t_0}| \geq \Delta \, \, .$$
	 In this way we consider only market fluctuations of amplitude 
	 $\Delta$.
	 Since the distribution of the returns is {\it almost} symmetric, 
	 the threshold $\Delta$ has been chosen equal for both positive and 
	 negative values.
	 Starting from $S_{t_1}$ we obtain with the same procedure
	 $S_{t_2}$.
   \item following the previous prescription we create a sequence of returns
	 $$\{r_{t_1,t_0}, r_{t_2,t_1}, \ldots, 
	    r_{t_k,t_{k-1}}, \ldots\}\,\, ,$$
	 from which we obtain the symbolic dynamics~:
	 \begin{equation}
	   c_k = \left \{ {\begin {array} {ccc}
				  -1 & {\mathrm if} & r_{t_k,t_{k-1}} < 0 \\
				  +1 & {\mathrm if} & r_{t_k,t_{k-1}} > 0
			   \end{array} } \right . \,\,\,\,.
	   \label{eq:filtrononlineare}
	 \end{equation}
	 We define $k$ as $\Delta$ {\it trading} time, i.e.
	 we enumerate only the transactions at which $\Delta$
	 is reached.
\end{itemize}

\begin{figure}[htb]
 \begin{center}
  \resizebox{0.7\textwidth}{!}{\includegraphics{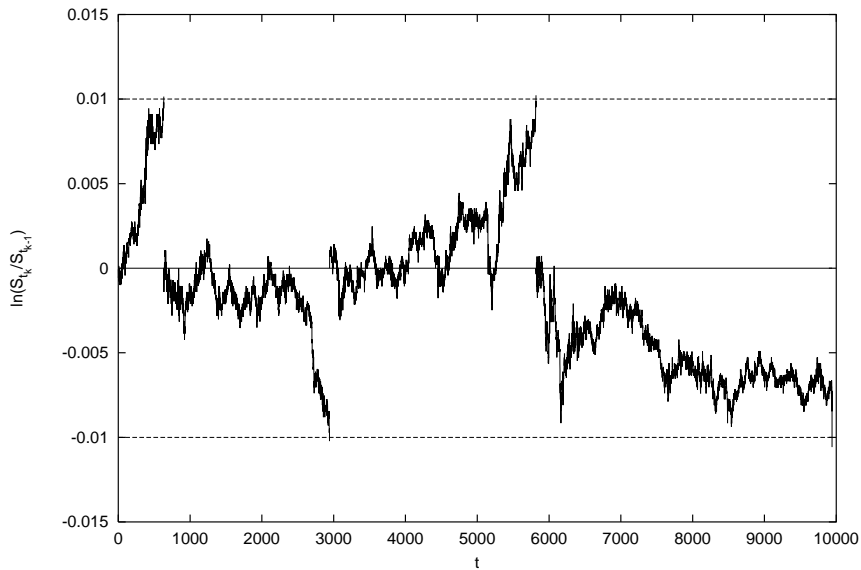}}
  \protect\caption{Evolution of $r_{t_k,t_{k-1}}$ with $\Delta=0.01$. $t_0=0$ corresponds
		   at 00:00:14 of October 1, 1992 in calendar time to the , 
		   and the $t_4=9939$ corresponds at 11:59:28 of October 2, 1992.}
  \label{fig:distrtd}
 \end{center}
\end{figure}

Let us notice that the variable $|r_{t_k,t_{k-1}}|$ has a narrow
distribution close to the threshold, and for all practical purposes
$|r_{t_k,t_{k-1}}|$ can be well approximated with $\Delta$.
In figure \ref{fig:distrtd} 
we show an example of evolution of the $r_{t_k,t_{k-1}}$.

\begin{figure}[htb]
 \begin{center}
  \resizebox{0.8\textwidth}{!}{\includegraphics{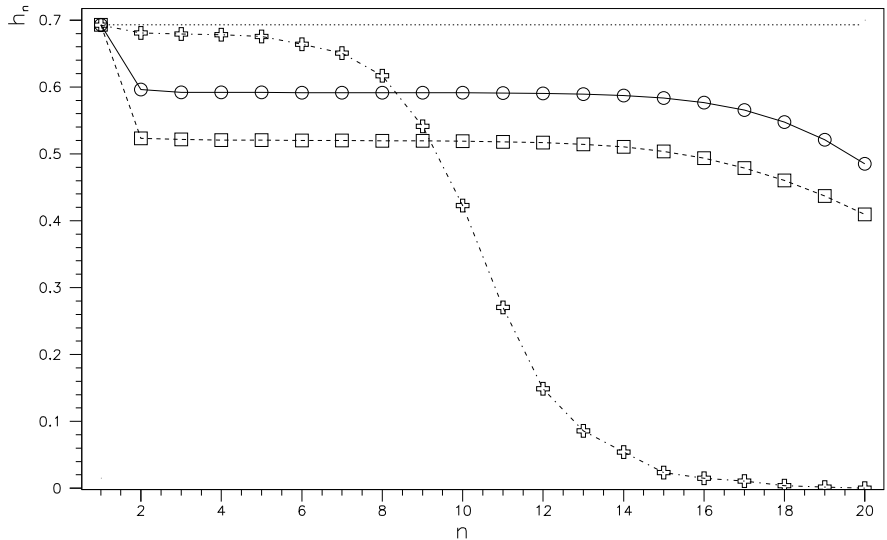}}
  \protect\caption{$h_n$ {\it versus} $n$. The three plots correspond 
		   to different value of $\Delta$~: $\Delta = 0.00005$ ($\circ$),
		   $\Delta = 0.0002$ ($\Box$) and $\Delta = 0.004$ ($+$). The
		   dotted line indicates $\ln(2)$.}
  \label{fig:hentrdelta}
 \end{center}
\end{figure}

The entropy analysis of the symbolic sequence $\{r_{t_k,t_{k-1}}\}$
gives a completely different result from the one in the previous section.
In figure \ref{fig:hentrdelta} it is shown that the entropy is clearly
different from $\ln(2)$ in a wide range of $\Delta$, i.e. there
is a set of $\Delta$ for which the {\it available} information 
(see eq. (\ref{eq:avainfo})) is very large.

\begin{figure}[htb]
 \begin{center}
  \resizebox{0.8\textwidth}{!}{\includegraphics{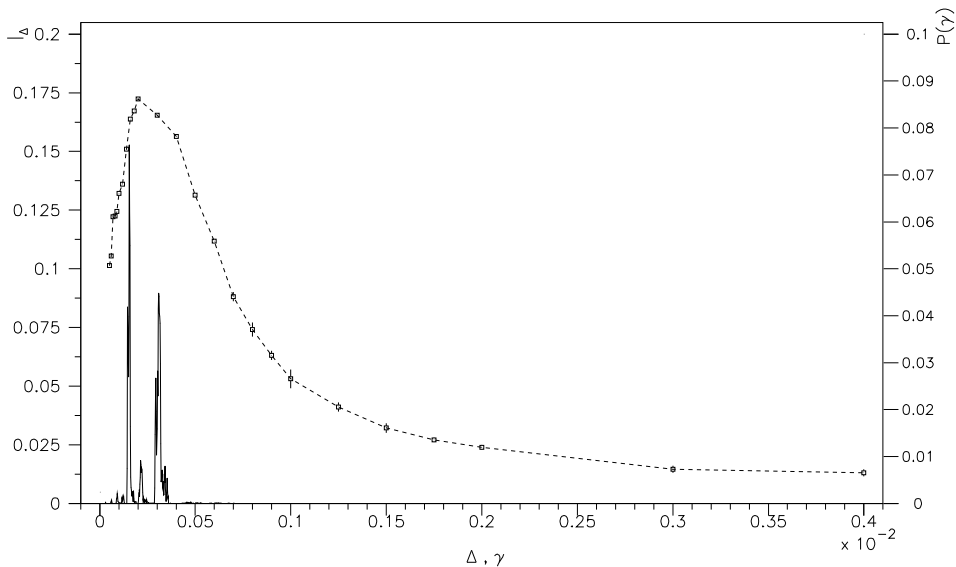}}
  \protect\caption{{\it Available} information $I_\Delta$ {\it versus} $\Delta$
	           (on the left), superimposed to the distribution of transaction 
		   costs, $P(\gamma)$ {\it versus} $\gamma$ (on the right).}
  \label{fig:entrtrans}
 \end{center}
\end{figure}

In figure \ref{fig:entrtrans} we plot the {\it available} 
information {\it versus} $\Delta$ and the distribution 
of transaction costs. 
Because these two quantities do not have similar size
they are plotted on different vertical scales but they are 
superimposed to make easier comparison between them. 
We observe that the maximum of the {\it available} 
information is almost in correspondence to the
maximum of the distribution of the transaction cost.

We have estimated the transaction costs $\gamma$ as
$$ \gamma_t = \frac{1}{2} \ln \frac{S_t^{(ask)}}{S_t^{(bid)}} \simeq 
	      \frac{S_t^{(ask)} - S_t^{(bid)}}{2S_t^{(bid)}}\,\,,$$
of course this is an upper bound for the true transaction cost.

We notice that the {\it available} information is almost
equal zero when we consider very small and very
large values of $\Delta$. 
These limit values cannot be reached for
two different reasons;
since $S_t$ can assume only discrete values,
it is not possible to take the limit $\Delta \rightarrow 0$.
In addition, we cannot compute $I\dd$ for large $\Delta$ because in
the sequence $r_{t_k,t_{k-1}}$ there are  not enough data
for an efficient statistical analysis.

\subsection{Profitable information}
We focus our attention on optimal strategies
in a financial market with non zero
{\it available} information.
We then show the economic relevance of such a quantity 
in case of {\it weak} efficiency of the market.

Consider the optimal growth rate strategy for a 
{\it patient} speculator.
The returns $\{r_{t_k,t_{k-1}}\}$ are almost symmetrically distributed 
and they can be well approximated
by the two threshold values $\Delta$ and $-\Delta$.

We shall only deal with the markovian case.
In fact as suggested by figure \ref{fig:hentrdelta} and shown 
in~\cite{BavVerVulI}, one has that the markovian approximation 
well reproduces the signal filtered with a fixed resolution $\Delta$.
The symmetry of the return distribution and the markovian nature 
of the process implies that the transition matrix is close to
\begin{equation}
\left( 
\begin{array}{cc}
p\dd     &  1-p\dd \\
1-p\dd   &  p\dd
\end{array}
\right) \,\,\,.
\label{eq:transition_matrix}
\end{equation}
where $p\dd$ is the probability to have $+1$ at 
$\Delta$ {\it trading} time $(t+1)$, knowing that $c_t$ was
$+1$ at time $t$.

In this particular case the {\it available} information is~:
\begin{equation}
{\it I\dd} = p\dd \ln (p\dd) 
	+ (1-p\dd) \ln (1-p\dd) + \ln (2)\,\,.
\label{eq:markov_information}
\end{equation}

We focus our attention on an investor who decides to diversify
his portfolio only in a security asset 
with a given interest rate return $r$,
and to invest, every $\Delta$ {\it trading} time $t$,
a fraction $l_t$ 
of his capital in the Deutschemark/US dollar exchange.
Our convention is that the fraction $l$ is positive
if he exchange dollars into marks, negative vice versa,
and we allow the speculator to borrow money from a bank.

We assume a vanishing interest rate return.
This hypothesis is reasonable:
in the period we are dealing with, the official discount rate  
fixed by the Federal Reserve is of 3 percent per year
and fluctuates between 5.75 and 8.75 percent in the German case. 
The {\it patient} speculator rehedges his portfolio on average 
every $66$ seconds when $\Delta$ is equal to the mean 
transaction cost.
The largest $\Delta$ corresponds to an average time of $8.6$ hours
of standby.
In the time scales involved the true interest rate return is 
about one hundred times smaller than $\Delta$:
the approximation of a vanishing interest rate appears to be fair.

We deal with the no transaction costs case:
this will allow us to understand easily the meaning of 
{\it available} information for a {\it patient} investor.
The more general situation with
transaction costs is treated in detail
in~\cite{BavVerVulII}.

Let us focus on the investment at time $t$:
the speculator commits a fraction $l_t$ in dollars.
At the following time step $t+1$ his capital becomes
\begin{equation}
W_{t+1}=(1+l_t c_{t+1} \Delta) W_t \,\,.
\label{eq:capital}
\end{equation}

We notice that, a consequence of the symmetry, is that the optimal
$l_t$ can assume only two values $l_t=c_t l$ where $l$ is a real number.

We define the {\it profitable} information as
the exponential rate of the capital of an investor 
who follows an optimal growth rate strategy.
The strict connection between this quantity and 
the {\it available} information was first noticed 
by Kelly~\cite{Kelly}, who, considering an elementary gambling game,
first gave an interpretation of Shannon entropy in the context
of optimal investment.

The computation of capital growth rate is a simple application of~\cite{Kelly}, 
and for the investment above described is
\begin{equation}
\lambda\dd(l) \equiv \lim_{T\to\infty}\frac{1}{T} \ln \frac{W_T}{W_0} 
= p\dd \ln (1+l \Delta) + (1-p\dd) \ln (1-l \Delta) \,\,.
\label{eq:lyapunov}
\end{equation}
It reaches its maximum for
\begin{equation}
l^* = \frac{2 p\dd -1}{\Delta} \,\,.
\label{eq:optimal_l}
\end{equation}
An intuitive consequence of equation (\ref{eq:optimal_l}) 
is that an anti-persistent return ($p\dd<1/2$),
as in the financial series we have considered,
implies 
that the optimal strategy
is to buy marks if the dollar rises,
and to do the opposite otherwise.
Of course a persistent case ($p\dd>1/2$) would imply
an $l_t$ greater than zero every time the positive
threshold $\Delta$ is reached.

From (\ref{eq:lyapunov}) and (\ref{eq:optimal_l}) one has that
the optimal growth rate 
is equal to the {\it available} information:
\begin{equation}
\lambda\dd^* =  \max_l \lambda\dd(l) 
 = p\dd \ln (p\dd) + (1-p\dd) \ln (1-p\dd) + \ln (2) = 
I\dd \,\,.
\label{eq:optimal_lyapunov}
\end{equation}

We stress that the equivalence between
{\it available} and {\it profitable} information,
if we forget the costs involved in this trading rule,
means that a speculator, 
who follows a particular strategy,
has the possibility to obtain a growth rate of his capital
exactly equal to this information:
this makes clear why we have called it {\it profitable}.

We underline that we have considered the growth rate
measuring the time in $\Delta$ {\it trading} time.
To obtain the exponential rate of the capital 
in the usual {\it calendar} time
we have to normalize (\ref{eq:lyapunov}) with the average exit 
time for the specified $\Delta$~\cite{LoretoServaVulpiani}.
For example for $\Delta = 0.0002$ corresponding to
the maximum {\it available} information
is characterized by
$\langle \tau \rangle = 66$ seconds.
This means that the average optimal growth rate is equal to 
$0.27$ percent per second. 

A naive consequence of previous results could be
that an efficient market hypothesis should be rejected.

Unfortunately (for the authors) this is not obvious.

We have previously noticed that the {\it available} information can be
transformed in {\it profitable},
let us now comment the
feasibility of the proposed trading rule.

When $\Delta$ is near the value of the maximum
{\it available} information,
the speculator changes his position with high frequency,
and $\Delta$ is comparable with transaction costs:
it is not any more possible to neglect them.

Furthermore in equation (\ref{eq:optimal_l}) $\Delta$ 
appears at the denominator,
and then the values of $l^*$ can be enormous.
For example for $\Delta$ corresponding to the maximum of the {\it available} information, 
the speculator who follows the optimal growth rate strategy, should
borrow $2830$ times the capital he has!
Even a small fluctuation from the expected average behavior
can lead to bankruptcy.

On the other hand if he wants to use reasonable values of $l^*$,
he has to chose a sufficiently large $\Delta$; 
in this situation the filtered series 
is almost indistinguishable from a ``random walk'' and then there is 
almost no {\it available} information.

We have now all the ingredients to comment
the shape of the {\it available} information
shown in figure~\ref{fig:entrtrans}.

The speculator cannot have a resolution
$\Delta$ lower than the transaction costs,
profits from such an investment would be in fact less than costs.
Therefore in this range of $\Delta$
the {\it available} information increases.
The discretization of the prize changes does not allow for 
reaching in a continuous way $\Delta=0$, where 
the ``random walk'' model is practically 
recovered as shown in the first part
of this section.

For $\Delta$ larger than the transaction costs the 
information can be exploited by proper strategies.
However,
small fluctuations are more difficult to detect and to distinguish 
from the ``noise'' and  
the {\it profitable} information is almost useless because of the
huge values of $l^*$ involved. 
This fact is even more evident when transaction costs are included.

On the other hand for large $\Delta$, the investors 
are able to discover the {\it available} 
information and to make it profitable with a feasible strategy. 
As a consequence, the efficient equilibrium is than restored 
for all practical purposes.

Let us briefly mention what happens instead to the {\it systematic}
investor. We can repeat exactly the above discussion and the
only difference is that now he decides to modify his portfolio
every $\tau$ {\it business} time. 
Because there is almost no {\it available} 
information (see subsection {\bf 3.1})
the optimal growth rate of his capital is vanishing even without
considering the costs involved in the transactions.

\section{Conclusions}

In this paper we have considered the long term anomalies
in the Deutschemark/US dollar
quotes in the period from October 1, 1992 to September 30, 1993 
and we have analyzed the consequences on the {\it weak} efficiency
of this market.

In section {\bf 2} we have shown the presence of long term
anomalies with two techniques: the structure functions and
a generalization of the usual correlation analysis.
In particular we have pointed out that ``random walk'' models
(or other self-affine models) cannot describe these features.

Once we have shown the existence of correlations in financial process, 
we have tested whether they allow for a {\it profitable} strategy.

With such a goal in mind, in section {\bf 3} we have first 
introduced a direct measure of the {\it available} information,
then we have shown in a particular case that this is equivalent 
to a {\it profitable} information.
In other words following a suitable trading rule
it is possible in absence of transaction costs
to have an exponential growth rate of the capital equal to this information.

We have measured the {\it available} information 
with a technique which reminds the Kolmogorov $\epsilon$ entropy. 
Two different codifications for financial series
(fixed lag $\tau$ and finite resolution $\Delta$) 
lead to completely different results.

The {\it available} information strongly depends on the kind of investment
the speculator has in mind.
We show that if he wants to change his position {\it systematically} 
at fixed lags $\tau$ the
{\it available} information is practically zero:
for this investor the market is efficient.

Instead, a {\it patient} investor, who 
waits to modify his portfolio till the asset has a fluctuation $\Delta$, 
observes a finite {\it available} information.

However, the existence of such a trading rule does not imply
that the investment is feasible in practice.
Namely we show that 
when reasonable investments are involved almost
no {\it available} information survives.
On the contrary, it is extremely difficult to use it when it is still present.

The technique described here
can be considered as a powerful tool to test {\it weak} efficiency~:
the speculator contributes to reach efficient equilibria
destroying the {\it available}
information that could be exploited in practice.
The efficiency hypothesis is then restored for almost all practical purposes.

\renewcommand{\theequation}{\Alph{appendixc}.\arabic{equation}}

\section*{Appendix}

In this appendix we show that 
if the correlations $C_q(\tau)$ exhibit a long range memory
$C_q(\tau) \sim \tau^{-\beta_q}$ then also
the variance $\delta_q(\tau)$ of the
{\it generalized cumulative absolute returns} $\{\chi_{t,q}(\tau)\}$
behaves at large $\tau$ as $\tau^{-\beta_q}$.

Making explicit expression of $\chi_{t,q}(\tau)$ (see equation
(\ref{eq:defchi})) one can write equation (\ref{eq:defvarchi}) as~:
$$
\delta_q(\tau) =
   \frac{1}{\tau^2} \sum_{\tau_1=0}^{\tau - 1} \sum_{\tau_2=0}^{\tau - 1}
      \avg{|r_{t+\tau_1}|^q |r_{t+\tau_2}|^q} - 
      \avg{|r_{t+\tau_1}|^q} \avg{|r_{t+\tau_2}|^q} \,\,.
$$
Taking into account the fact that $r_t$ is a stationary process, 
and using the definition of $C_q(\tau)$, one has:
$$
\delta_q(\tau) =
{1\over \tau} C_q(0)
+ {2\over \tau^2} \sum_{\tau > \tau_1>\tau_2 \ge 0} C_q(\tau_1-\tau_2) 
$$
where
$$ C_q(0)= \avg{|r_t|^{2q}} - \avg{|r_t|^q}^2 \,\,. $$
The expression of $\delta_q(\tau)$  can be rewritten as:
$$ 
\delta_q(\tau) = {1\over \tau} C_q(0)
+ {2\over \tau^2} \sum_{\tau_1=1}^{\tau-1} (\tau-\tau_1)\  C_q(\tau_1) \,\,.
$$
Under the hypothesis $C_q(\tau) \sim \tau^{-\beta_q}$, 
one has for large $\tau$
$$ 
{2\over \tau^2} \sum_{\tau_1=1}^{\tau-1} (\tau-\tau_1) \ C_q(\tau_1) 
   \sim \tau^{-\beta_q} \,\,,
$$
which leads to~:
$$ \delta_q(\tau) = O(\tau^{-1}) + O(\tau^{-\beta_q}) \,\,.$$
Since $\beta_q \le 1$, the thesis follows, i.e.~:
$$ \delta_q(\tau) \sim \tau^{-\beta_q} \,\,.$$

\section*{Acknowledgment}

We thank Luca Biferale and Rosario Mantegna for many useful discussions
at an early stage of this research. 
We also thank Paolo Muratore Ginanneschi and Alberto Bigazzi for a
careful reading of the manuscript.

\end{document}